\documentclass[prd,showpacs]{revtex4}
\usepackage{graphicx}
\usepackage{amsmath}
\usepackage{amssymb}
\begin{document}

\renewcommand{\theequation}{\thesection.\arabic{equation}}

\newcommand{\re}{\mathop{\mathrm{Re}}}

\newcommand{\be}{\begin{equation}}
\newcommand{\ee}{\end{equation}}
\newcommand{\bea}{\begin{eqnarray}}
\newcommand{\eea}{\end{eqnarray}}

\title{Brane $f(R)$ gravity cosmologies}

\author{Adam Balcerzak}
\email{abalcerz@wmf.univ.szczecin.pl}
\affiliation{\it Institute of Physics, University of Szczecin, Wielkopolska 15,
          70-451 Szczecin, Poland}

\author{Mariusz P. D\c{a}browski}
\email{mpdabfz@wmf.univ.szczecin.pl}
\affiliation{\it Institute of Physics, University of Szczecin, Wielkopolska 15,
          70-451 Szczecin, Poland}

\date{\today}

\input epsf

\begin{abstract}

By the application of the generalized Israel junction conditions we derive cosmological equations
for the fourth-order $f(R)$ brane gravity and study their
cosmological solutions. We show that there exists a non-static solution which describes a
four-dimensional de-Sitter $(dS_4)$ brane embedded in a five-dimensional anti-de-Sitter $(AdS_5)$ bulk
for a vanishing Weyl tensor contribution. On the other hand, for the case of a non-vanishing Weyl tensor
contribution, there exists a static brane solution only. We claim that in order to get some more general
non-static $f(R)$ brane configurations, one needs to admit a dynamical matter energy-momentum tensor
in the bulk rather than just a bulk cosmological constant.

\end{abstract}


\maketitle

\section{Introduction}

Similarly as in the standard general relativity, it is interesting to consider generalizations of the brane
universes \cite{RS,brane1,brane2}. Some of such generalizations are the higher-order brane gravity theories of which the simplest is
$f(R)$ gravity \cite{f(R)} (for the most recent reviews see Refs. \cite{faraoni10,NOreview}). However, the combination of brane models with higher-order theories is non-trivial,
since, except for the Lovelock (or, in the lowest-order, the Gauss-Bonnet) densities \cite{GBbrane,meissner01}, one faces ambiguities of the
quadratic delta function contributions to the field equations. This problem was first challenged succesfully
in our earlier works \cite{paper1,paper2}, in which we found the ways to avoid ambiguities not only for $f(R)$
brane theories (see e.g. \cite{borzeszkowski, hinterb}), but also for more general actions which depend arbitrarily on the three curvature invariants
$f(R, R_{ab}R^{ab}, R_{abcd}R^{abcd})$, although the linear combination of these invariants was studied in Ref. \cite{NOlinear}.
One of the methods applied, was the reduction of the fourth-order brane gravity
to the second-order theory by introducing an extra degree of freedom -- the scalaron \cite{paper1,paper2}.
Such a procedure leads to the second-order gravity which is just the scalar-tensor Brans-Dicke
gravity \cite{bd} with a Brans-Dicke parameter $\omega =0$, and an appropriate scalaron potential
(with the scalaron playing the role of the Brans-Dicke field). We then obtained the Israel junction conditions \cite{israel66}
which generalized both the conditions obtained in Ref. \cite{BDbrane,BDbranenoZ2} for the Brans-Dicke
field without a scalar field potential, and also the conditions derived in Ref. \cite{deeg,deruelle07}
for $f(R)$ brane gravity. The junction conditions which did not assume scalaron continuity,
but for a static brane, were presented in Ref. \cite{afonso}.

In this paper we apply the Israel junction conditions for $f(R)$ Friedmann-Robertson-Walker metric brane configurations
and study the set of their admissible cosmological solutions. In Section II we present $f(R)$ brane models and derive the set of field
equations. In Section III we apply the field equations to cosmology. In Section IV we give our conclusions.

\section{$f(R)$ gravity on the brane}
\label{fRgrav}
\setcounter{equation}{0}

Let us consider the $f(R)$ gravity on the brane described by the action \cite{paper1}
\bea
\label{ac}
S_{p} &=& \frac{1}{2\kappa_5^2} \int_{M_p} d^5 x \sqrt{-g} f(R)+ S_{bulk,p}~,
\eea
where $R$ is the Ricci scalar, $\kappa_5^2$ is a five-dimensional Einstein constant, $S_{bulk,p}$ is the bulk
matter action
($p=1,2$), $M_p$ is the spacetime volume. The action (\ref{ac}) gives fourth-order field equations.
It is then advisable to use an
equivalent action
\begin{eqnarray}
\label{equiv2}
\bar{S}_{p} &=& \int_{M_p}d^5 x\sqrt{-g}\{f'(Q)(R-Q) + f(Q)\}
\end{eqnarray}
where $Q$ is an extra field (Lagrange multiplier), and $f'(Q) =
df(Q)/dQ$. The equation of motion which comes from (\ref{equiv2}) is
just $Q=R$, provided that $f''(Q) \neq 0$, so that $f'(Q)$ may be interpreted as an extra scalar
field (called the scalaron)
\be
\label{scalaron}
\phi=f'(Q) = f'(R)~,
\ee
and the action can be rewritten as
\begin{eqnarray}
\label{equiv3}
\bar{S}_{p} &=& \int_{M_p} d^5 x \sqrt{-g}\{\phi R - V(\phi)\} + S_{bulk,p} .
\end{eqnarray}
where $V(\phi) = -\phi R(\phi) + f(R(\phi))$ \cite{paper2}. The action (\ref{equiv3})
is equivalent to a scalar-tensor Brans-Dicke gravity with a Brans-Dicke parameter $\omega=0$.
One of the ways to derive the junction conditions for the theory described by the action (\ref{equiv3})
is to append it with an appropriate Hawking-Lutrell boundary term which reads as \cite{lutrell}
\begin{equation}
\label{HL}
S_{HL_{p}}=-2(-1)^{p}\epsilon \int_{\partial M_p} \sqrt{-h} \phi K d^4x~,
\end{equation}
where $K$ is the trace of the extrinsic curvature tensor $K_{ab}$, $h$ is the determinant of the induced metric
$h_{ab} = g_{ab} - \epsilon n_an_b$, $n^a$ is a unit normal vector to a boundary $\partial M_p$, and
$\epsilon = 1$ $(\epsilon = -1)$ for a timelike (a spacelike) brane, respectively.
The total action of the theory is then
\begin{eqnarray}
\label{equiv4}
\bar{S}_{tot_{p}} &=& \bar{S}_{p}+S_{LH_{p}}.
\end{eqnarray}
The variation of the action (\ref{equiv4}) leads to the following junction conditions for $f(R)$ gravity
theory \cite{paper2} ($a, b, \ldots = 0, 1, 2, 3, 5$)
\begin{eqnarray}
\label{jc2}
[K]&=& 0~, \\
\label{jc21}
S^{ab}n_{a}n_{b}&=& 0~, \\
\label{jc22}
S^{ab}h_{ac}n_{b}&=& 0~, \\
\label{jc24}
-h_{ab}[\phi_{,c}n^{c}]-[\phi]Kh_{ab} + [\phi K_{ab}]
&=&
\epsilon \kappa_5^2 S^{cd}h_{ca}h_{db},
\end{eqnarray}
where for an arbitrary quantity $A$ we have defined a discontinuity (a jump) of $A$ as: $[A]\equiv A^+ - A^-~$.
Here $S_{ab}$ is the brane energy-momentum tensor (cf. later). In particular, the condition (\ref{jc2}) comes
from the requirement that the variation of the Hawking-Lutrell boundary term (\ref{HL}) should vanish.

These junction conditions can be compared with previously obtained in Ref. \cite{deeg} (see their Eqs. (12)
and (13))
and in Ref. \cite{deruelle07} (see their Eq. (3.11)). The difference is that we have not assumed the continuity
of the scalaron on the brane.  If we do so, i.e. assume that $[\phi]=0$ which due to the definition
(\ref{scalaron})
implies $[R]=0$, and additionally impose the mirror
symmetry $g_{ab}=g_{ab}(|n|)$, where $n$ is a normal Gaussian coordinate originating at the brane, then
the junction conditions (\ref{jc2})-(\ref{jc24}) take the form
\begin{eqnarray}
\label{jc3a}
K&=&0~, \\
\label{jc3b}
\phi_{,c}n^c &=&- {\kappa_5^2 \over 8} S^{ab}h_{ab}~,\\
\label{jc3c}
 K_{ab}&=& {\kappa_5^2 \over  2 \phi}\{S^{cd}h_{ca}h_{db}-{h_{ab} \over
4}S^{cd}h_{cd}\}~.
\end{eqnarray}

We can now express the equations above in terms of the Ricci scalar $R$ instead of the scalaron $\phi$.
Using the Gaussian coordinate system we obtain
\begin{eqnarray}
\label{jc3aa2}
R&=&0~, \\
\label{jc3a2}
K&=&0~, \\
\label{jc3b2}
f''(R) R_{,n} &=&- {\kappa_5^2 \over 8} \widetilde{S}~,\\
\label{jc3c2}
f'(R) \widetilde{K}_{ab}&=& {\kappa_5^2 \over  2}\widetilde{S}^{ab}~,
\end{eqnarray}
where we have used the definition of a traceless part of the brane energy-momentum tensor
$\widetilde{S}^{ab}=S^{cd}h_{ca}h_{db}-(1/4)h_{ab}S^{cd}h_{cd}$, the definition of
the traceless part of the extrinsic curvature tensor
$\widetilde{K}^{ab}=K^{cd}h_{ca}h_{db}-(1/4)h_{ab}K^{cd}h_{cd}$  and the definition of the trace
of the brane energy-momentum tensor $\widetilde{S}=S^{ab}h_{ab}$. The condition (\ref{jc3aa2}) is a consequence
of the definition (\ref{scalaron}). The junction conditions  (\ref{jc3aa2})-(\ref{jc3c2}) coincide with those
obtained in Ref. \cite{deruelle07} (Eq. (3.11)).

Now we can apply the junction conditions (\ref{jc2})-(\ref{jc24}) for $f(R)$ gravity in order to obtain the effective
Einstein equations on the brane. The manipulation of the Gauss-Codazzi equation \cite{wald}
\begin{eqnarray}
\label{coddcov}
{}^{(5)}R_{cd}h^{d}_{~a}n^{c}= -D_{c}K^{c}_{~a} + D_{a}K
\end{eqnarray}
where $D_a$ means a 4-dimensional covariant derivative on the brane, leads to a standard decomposition of a
four-dimensional Einstein tensor $^{(4)}G_{ab}$ in the form \cite{gronhervik}
\begin{eqnarray}
\label{Gdekomp}
\nonumber {}^{(4)}G_{ab}
=  KK_{ab}
-K^{~c}_{a}K_{bc} -{1 \over 2}h_{ab}  (K^2
-K^{cd}K_{cd}) - {}^{(5)}E_{ab}\ \\
 + {2\over3}[{}^{(5)}G_{cd}h^{c}_{~a}h^{d}_{~b}+ h_{ab}({}^{(5)}G_{cd}n^{c}n^{d} - {1\over4}{}^{(5)}G)],
\end{eqnarray}
where $E_{ab}$ is an electric part of the bulk Weyl tensor projected onto the brane. We also assume that in the neighborhood of the brane, the normal vector field $n^{a}$ fulfills the geodesic
equations $n^{a}\nabla_{a}n^{b}=0$ (geodesic gauge). Using this last assumption, the following relations
are derived (see the Appendix)
\begin{eqnarray}
\label{boxdekomp}
{}^{(5)} \Box \phi &=& {}^{(4)} \Box \phi + K n^{a}\nabla_{a}\phi + (n^{a}\nabla_{a})^{2}\phi ,\\
\label{boxdekomp2}
h^{c}_{~a}h^{d}_{~b}\nabla_{c}\nabla_{d}\phi &=& D_{a} D_{b}\phi + K_{ab}n^{c}\nabla_{c}\phi.
\end{eqnarray}
Here $\nabla_{c}$ means the 5-dimensional (bulk) covariant derivative. Assuming that the matter in the bulk has the form of the 5-dimensional cosmological constant
$T_{ab}= -g_{ab} {^{(5)}\Lambda}$,
the variation of the action (\ref{equiv4}) gives the  following field equations in the bulk
\begin{eqnarray}
\label{pierbulk}
{}^{(5)}G_{ab}&=&-{1\over 2\phi} g_{ab}V(\phi) + {1\over \phi} g_{ab} {}^{(5)}
\Box \phi - {1\over \phi} \phi_{;ab} + {\kappa_5^2 \over \phi} g_{ab}{^{(5)}\Lambda}, \\
\label{2eqmot}
{}^{(5)}R&=&- {\partial V(\phi) \over \partial \phi}\equiv - W(\phi).
\end{eqnarray}
Substituting  (\ref{jc3a})-(\ref{jc3c}), (\ref{boxdekomp}), (\ref{boxdekomp2}), and (\ref{pierbulk})
to (\ref{Gdekomp}), one obtains the effective Einstein equations on the brane as
\begin{eqnarray}
\label{GenEin}
 {}^{(4)}G_{ab}&=& \left({\kappa_5^2 \over 2 \phi}\right)^{2}Q_{ab}-{1 \over4} h_{ab}{V(\phi)\over \phi}
  + {2\over3}{1\over \phi}h_{ab} {{}^{(4)} \Box \phi } \nonumber
 \\  &-&{2\over3}{1\over \phi} D_{a}D_{b}\phi + {\kappa_5^2 \over 2\phi}{^{(5)}\Lambda} h_{ab} -{}^{(5)} E_{ab},
\end{eqnarray}
where
\begin{eqnarray}
Q_{ab}=-{2\over 3}\lambda(\tilde{T}_{ab} -{1\over 4} \tilde{T} h_{ab}) + {2\over 3}\tilde{T}\tilde{T}_{ab}
-\tilde{T}^{c}_{~b} \tilde{T}_{ca} +{1\over2}h_{ab}(\tilde{T}_{cd}\tilde{T}^{cd} - {11\over24}\tilde{T}^2)
\end{eqnarray}
and
\begin{eqnarray}
\tilde{T}_{ab}:= \lambda h_{ab}+ S_{ab}.
\end{eqnarray}
This should be appended by the conservation law for the matter energy-momentum tensor on the brane
(see Appendix \ref{appendixB})
\begin{equation}
\label{con}
D_a S^{a}_{~b}=0~.
\end{equation}

\section{$f(R)$ Friedmann cosmology on the brane}
\label{fRcosm}
\setcounter{equation}{0}

We restrict ourselves to the case of the matter in the bulk in the form of the cosmological constant.
This allows to assume that the bulk spacetime is an Einstein space
\begin{equation}
\label{bulki}
{^{(5)}} G_{ab}=-{^{(5)}\Lambda_{eff}} g_{ab},
\end{equation}
where ${^{(5)}\Lambda_{eff}}>0$ is an effective cosmological constant.
The five-dimensional line element reads as
 \begin{eqnarray}
 \label{fred}
ds^{2}=-b^{2}(n,t)dt^{2} + a^{2}(n,t)\left[ \frac{dr^{2}}{1-kr^{2}}+r^2d\Omega^{2}\right]
+ dn^{2}~,
\end{eqnarray}
where $k=0, \pm 1$. The electric part of Weyl tensor $E_{ab}$ can be expressed in the following form \cite{gronhervik}
\begin{eqnarray}
{}^{(5)} E_{ab}= \mathcal{F}[u_{a}u_{b} + {1\over3}(h_{ab}+u_{a}u_{b})].
\end{eqnarray}
In the case with vanishing $\mathcal{F}$, we deal with a non-static Friedmann-Robertson-Walker brane
(\ref{fred}) embedded in an $AdS_{5}$ bulk \cite{mannheim}. Moreover, the junction condition
(\ref{jc3b}) requires that the trace of the brane energy-momentum tensor
vanishes $S=\tilde{T}+4\lambda=0$ (the second bulk equation (\ref{2eqmot}) sets the scalaron to be
a constant because of the constancy of the curvature in the bulk). The assumption that the energy-momentum
tensor of the matter on the brane is a perfect fluid
\be
S_{ab} = (\rho + p) u_a u_b + p g_{ab}~~,
\ee
which fulfills the barotropic equation of state $p = w \rho$ with the four-velocity vector $u^a = \delta^a_0$,
gives the effective $f(R)$ gravity Friedmann equations on the brane in the form
\begin{eqnarray}
\label{GenEin1}
 -\mathcal{F}- {\kappa_5^2 \over 2\phi}{^{(5)}}\Lambda +{V(\phi)\over 4\phi} &-&
3 {{k+\dot{a}^2}\over a^2} + 2{\dot{a}\over a}{\dot{\phi}\over \phi}
\\ \nonumber &=&-\left({\kappa_5^2 \over 8\phi}\right)^2 (p+\rho)(9p+ \rho -8\lambda), \\
\label{GenEin2}
{\mathcal{F}\over 3} -{\kappa_5^2 \over 2\phi}{^{(5)}}\Lambda + {V(\phi)\over 4\phi}-
{{k+\dot{a}^2}\over a^2}+ {4\over3}{\dot{a}\over a}{\dot{\phi}\over \phi}&-& 2{\ddot{a}
\over a}+{2\over3}{\ddot{\phi} \over \phi}
\\ \nonumber &=& {1\over 3} \left({\kappa_5^2 \over 8 \phi}\right)^2(p+\rho)(21p+ 13\rho -8\lambda).
\end{eqnarray}
Now the brane energy-momentum tensor conservation law (\ref{con})
\begin{eqnarray}
\label{concon}
\frac{\dot{\rho}}{\rho} = -3(w+1) \frac{\dot{a}}{a}
\end{eqnarray}
can be integrated to give
\begin{eqnarray}
\label{solrho}
\rho&=&\rho_{0}a^{-3(w+1)}
\end{eqnarray}
The requirement that the trace of the brane energy-momentum tensor vanishes imposes a condition 
that the energy density of the matter on the brane is constant, i.e.,
\begin{equation}
\label{roconst}
\rho = \rho_{0}=-\frac{4\lambda}{(1+3w)} = {\rm const.}
\end{equation}
Multiplying (\ref{GenEin2}) by three, and adding it to (\ref{GenEin1}), we get one cosmological
equation to solve (for simplicity we consider flat $k=0$ models only)
\begin{equation}
\label{cosmo1}
-2\kappa_5^2 \frac{\Lambda}{\phi} + \frac{V(\phi)}{\phi} +
6 \frac{\dot{a}}{a}\frac{\dot{\phi}}{\phi}+
2\frac{\ddot{\phi}}{\phi}-6\left(\frac{\dot{a}^2}{a^2}
+ \frac{\ddot{a}}{a}\right)=\frac{3}{4}{\left(\frac{\kappa_5^2}{2}
\frac{\rho}{\phi}\right)^2}(w+1)^2~~.
\end{equation}
The Eq. (\ref{2eqmot}) forces the scalaron $\phi$ to be constant $\phi=\phi_{0}$, as well.
On the other hand, the eq. (\ref{pierbulk}) leads to the following relation
\begin{equation}
\label{poten}
\frac{V(\phi_{0})}{2\phi_{0}} - \frac{\kappa_5^2 {^{(5)}}\Lambda}{\phi_{0}} = {^{(5)}\Lambda_{eff}} .
\end{equation}
The Eq. (\ref{poten}) shows that in the case of a constant scalaron, the term ${^{(5)}\Lambda_{eff}}$ plays
the role of a five-dimensional effective cosmological constant in the bulk.
We can independently fix the value of the $\phi_{0}$ by a  choice of the shape of the function 
$W(\phi)$ near the brane using (\ref{2eqmot}) as 
\begin{equation}
\label{vi}
\phi_{0}=W^{-1}\left(-\frac{10}{3}{^{(5)}\Lambda_{eff}}\right).
\end{equation}
Now for a fixed value of $V(\phi_{0}) \equiv V_0$, Eq. (\ref{poten}) gives
\begin{eqnarray}
\label{ka2}
{^{(5)}\Lambda_{eff}}= \frac{V_0- 2\kappa_5^2 {^{(5)}}\Lambda}{2\phi_0}.
\end{eqnarray}
Combining (\ref{cosmo1}) with (\ref{poten}), and assuming that $w=-1$, we obtain
\begin{equation}
\label{cosmo2}
3\left(\frac{\dot{a}^2}{a^2} + \frac{\ddot{a}}{a}\right)= {^{(5)}\Lambda_{eff}}.
\end{equation}
The non-static solution of (\ref{cosmo2}) takes the form
\begin{equation}
\label{sol1}
a=\tilde{a}_{0} \exp{(H_0 t)}~,
\end{equation}
and it is consistent with the solution of (\ref{GenEin1}) for $\mathcal{F}=0$. For $w\neq-1$, the continuity
equation (\ref{concon}) implies a constant scale factor $a(t)=a_{0}$, and the generalized Friedmann
eqs. (\ref{GenEin1}) and (\ref{GenEin2}) become inconsistent. The solution (\ref{sol1}) describes an
embedding of a de-Sitter $(dS_4)$ brane in an anti-de-Sitter $(AdS_{5})$ bulk provided that
\begin{eqnarray}
\label{ft}
H_{0}=\sqrt{\frac{{^{(5)}\Lambda_{eff}}}{6}}=\sqrt{\frac{V_0-2\kappa_5^2 {^{(5)}}\Lambda}{12\phi_0}}
\end{eqnarray}
(note that because of the assumption that ${^{(5)}\Lambda_{eff}}>0$, we need
$V_0 > 2\kappa_5^2 {^{(5)}}\Lambda$).
The Eq. (\ref{ft}) is a fine-tuning condition for the value of the bulk cosmological
constant ${^{(5)}}\Lambda$
and the potential $V(\phi)$, which is responsible for the value of $\phi_0$ and $V_0$.
The special case with
\begin{eqnarray}
\label{ft2}
V_0=2\kappa_5^2 {^{(5)}}\Lambda
\end{eqnarray}
gives $H_{0}=0$, and the solution (\ref{sol1}) describes a static Minkowski brane which is a flat analogue of
the Einstein Static Universe. In fact, the condition (\ref{ft2}) is a special case of the fine-tuning 
relation (\ref{ft}), and can be 
interpreted as a necessary and a sufficient condition for the existence of a static brane in the model
with $w=-1$ and a vanishing Weyl tensor contribution $\mathcal{F}=0$.

If the Weyl tensor contribution is non-vanishing, i.e., if $\mathcal{F} \neq 0$, it is then possible to embed
a static Friedmann-Robertson-Walker brane (\ref{fred}) in a bulk with the cosmological constant $ {^{(5)}}\Lambda$
only \cite{mannheim}. In such a case, the solution of (\ref{bulki}) for the metric
(\ref{fred}) has the form \cite{mannheim}
\begin{eqnarray}
\label{bulksol}
a^2(n)=f(n),~{\rm where}~~~~ f(n)=\gamma e^{2H_0|n|}+\delta e^{-2H_0|n|}~,\\
b^2(n)=\frac{e^{2}(n)}{f(n)},~{\rm where}~~~~ e(n)=\gamma e^{2H_0|n|}-\delta e^{-2H_0|n|}~,
\end{eqnarray}
with the brane at $n=0$. Using the transformation of the metric components 
\begin{equation} 
f(n) \rightarrow  \frac{f(n)}{\gamma+\delta} \hspace{0.2cm} {\rm and} \hspace{0.2cm} 
\frac{e^{2}(n)}{f(n)} \rightarrow \frac{e^{2}(n)}{f(n)} \frac{\gamma+\delta}{(\gamma-\delta)^2}~,
\end{equation}
which is equivalent to a rescaling of the coordinates, we obtain a Minkowski brane (for $n=0$). We
then compute the non-vanishing components of the electric part of the Weyl tensor $E^{a}_{~b}$ and 
the corresponding term $\mathcal{F}$ at the brane ($n=0$) as
\begin{eqnarray}
\label{Weil}
E^{1}_{~1}=E^{2}_{~2}=E^{3}_{~3}=-E^{0}_{~0}=\mathcal{F}=\frac{2\gamma\delta}{(\gamma+\delta)^2}{^{(5)}}
\Lambda_{eff}~.
\end{eqnarray}

For $w\neq -1$, the solution of Eqs. (\ref{concon}), (\ref{GenEin1}) and (\ref{GenEin2}) gives
\begin{eqnarray}
\label{EinSol}
w&=&\frac{1-3M\pm 2\sqrt{M}}{9M-1}, ~~~~{\rm where}~ M=\frac{\phi_0(V_0-2\kappa_5^2\Lambda)}{3(\kappa_5^2\lambda)^{2}} > 0\\
\label{ft3}
\mathcal{F}&=&\frac{3}{2}\left(\frac{\kappa_5^2\lambda}{\phi_0}\right)^{2}\sqrt{M}.
\end{eqnarray}
Note that (\ref{EinSol}) and (\ref{ft3}) requires that $\phi_{0}>0$.
The solution above describes a flat static Minkowski brane which is a flat analogue of
the Einstein Static Universe. In fact, from (\ref{Weil}) and (\ref{ft3})
we have
 \begin{eqnarray}
 \label{ft4}
\frac{2\gamma\delta}{(\gamma+\delta)^2}{^{(5)}}\Lambda_{eff}=
\frac{3}{2}\left(\frac{\kappa_5^2\lambda}{\phi_0}\right)^{2}\sqrt{M}.
 \end{eqnarray}
The condition (\ref{ft4}) means that to support a flat static Minkowski brane one needs to
fine-tuned the values of the parameters  $\phi_0$, $V_0$,  ${^{(5)}}\Lambda$, $\lambda$, $\gamma$ and $\delta$.


\section{Conclusions}

In this paper we have studied brane universes within the framework of the fourth-order $f(R)$ gravity theory.
We applied the junction conditions obtained in our earlier papers \cite{paper1,paper2} in order to get the
set of the field equations which were applicable to cosmology. We conclude that for the matter with
a barotropic equation of state $p=w\rho$ on the non-static Friedmann-Robertson-Walker brane (\ref{fred})
embedded in a five-dimensional anti-de-Sitter $(AdS_{5})$ bulk (with vanishing Weyl tensor contribution
$\mathcal{F}=0$), and the matter in the bulk having the form of
a cosmological constant ${^{(5)}}\Lambda$, there is only one case with $w=-1$ that possesses the solution in the
form of the exponential evolution (\ref{sol1}) which is a four-dimensional de-Sitter $(dS_4)$ brane embedded in
a five-dimensional anti-de-Sitter $(AdS_5)$ bulk. The case with the Friedmann-Robertson-Walker brane (\ref{fred})
embedded in a bulk with the cosmological constant ${^{(5)}}\Lambda$ and non-vanishing Weyl tensor contribution
$\mathcal{F}\neq 0$ allows the solution in the form of the  flat static Minkowski  universe (\ref{EinSol}) only.
The cosmological constant ${^{(5)}}\Lambda$ in the bulk implies the constant curvature and whence the vanishing
of the trace of the brane energy-momentum tensor $S$. This is an extremely strong condition that makes
the energy density constant. In conclusion, we claim that more non-static configurations on the brane
are possible if we assume a dynamical matter energy-momentum tensor in the bulk.


\section{Acknowledgements}

We acknowledge the support of the Polish Ministry of
Science and Higher Education grant No N N202 1912 34 (years 2008-10). We are indebted to Bogus{\l }aw Broda,
Krzysztof Meissner, Sergei Odintsov and Yuri Shtanov for useful discussions.

\appendix

\section{Derivation of useful geometric formulas from Section \ref{fRgrav}.}
\label{do_wody2}

The formula (\ref{boxdekomp}) can be obtained as follows
\bea
^{(5)}\Box \phi &=&g^{ab}\nabla_{a}\nabla_{b}\phi=h^{ab}\nabla_{a}\nabla_{b}\phi
+n^{a}n^{b}\nabla_{a}\nabla_{b}\phi \\ \nonumber
&=&h^{ab}D_{a}D_{b}\phi+h^{ab}K_{ab}(n^c \nabla_c) \phi +
(n^c \nabla_c)^2 \phi =^{(4)}\Box \phi + K (n^c \nabla_c) \phi + (n^c \nabla_c)^2 \phi.
\eea
where we have used
\bea
n^{c}n^{d}\nabla_{c}\nabla_{d}\phi=n^{c}\nabla_{c}(n^{d}\nabla_{d}H)
-n^{c}(\nabla_{c}n^{d})(\nabla_{d}\phi)=n^{c}\nabla_{c}(n^{d}
\nabla_{d}\phi)=(n^c \nabla_c)^2 \phi.
\eea

The formula (\ref{boxdekomp2}) can be obtained as follows
\bea
h^{c}_{~a}h^{d}_{~b}\nabla_{c}\nabla_{d}\phi&=&
h^{c}_{~a}h^{d}_{~b}\nabla_{c}(g^{e}_{~d}\nabla_{e}\phi)=
h^{c}_{~a}h^{d}_{~b}\nabla_{c}[(h^{e}_{~d}+n^{e}n_{d})\nabla_{e}\phi]=\\ \nonumber
&=&h^{c}_{~a}h^{d}_{~b}\nabla_{c}(h^{e}_{~d}\nabla_{e}\phi)+
h^{c}_{~a}h^{d}_{~b}(\nabla_{c}n^{e})n_{d}(\nabla_{e}\phi)+\\ \nonumber
&+&h^{c}_{~a}h^{d}_{~b}(\nabla_{c}n_{d})(n^{e}\nabla_{e}\phi)+\\ \nonumber
&+&h^{c}_{~a}h^{d}_{~b}n^{e}n_{d}\nabla_{c}\nabla_{e}\phi=\\ \nonumber
&=&D_{a}(h^{e}_{~b}\nabla_{e}\phi)+
h^{c}_{~a}h^{d}_{~b}\nabla_{c}n_{d}(n^{e}\nabla_{e}\phi)=\\ \nonumber
&=&D_{a}D_{b}\phi+K_{ad}h^{d}_{~b}(n^c \nabla_c) \phi=
D_{a}D_{b}\phi+\\ \nonumber
&+&K_{ab}(n^c \nabla_c) \phi.
\eea

It is also useful to prove that
\bea
\label{deko}
h^{d}_{~a}n^{c}\nabla_{d}\nabla_{c}\phi&=&
h^{d}_{~a}\nabla_{d}(n^{c}\nabla_{c}\phi)-h^{d}_{~a}(\nabla_{d}n^{c})
(\nabla_{c}\phi)=\\ \nonumber
&=&h^{d}_{~a}\nabla_{d}(n^c \nabla_c) \phi -
K_{a}^{~c}\nabla_{c}\phi=D_{a}(n^c \nabla_c )\phi-
K_{a}^{~e}g_{e}^{~c}\nabla_{c}\phi=\\ \nonumber
&=&D_{a}(n^c \nabla_c) \phi-K_{a}^{~e}(h_{e}^{~c}+
n_{e}n^{c})\nabla_{c}\phi=\\ \nonumber
&=&D_{a}(n^c \nabla_c)\phi -K_{a}^{~e}h_{e}^{~c}\nabla_{c}\phi-
K_{a}^{~e}n_{e}n^{c}\nabla_{c}\phi\\ \nonumber
&=&D_{a}(n^c \nabla_c \phi)-K_{a}^{~e}D_{e}\phi.
\eea

\section{Derivation of the brane energy-momentum tensor conservation}
\label{appendixB}

The  conservation law for the brane energy-momentum tensor can be obtained as follows. We take the covariant
derivative of the left hand side of the equation (\ref{jc3c}) multiplied by $\phi$ and get
\begin{eqnarray}
\label{DS1}
D_{a}(\phi K^{a}_{~b})&=& K^{a}_{~b} D_{a}\phi + \phi D_{a} K^{a}_{~b}.
\end{eqnarray}
Next, we use the Gauss-Coddazzi equation \ref{coddcov} together with the condition (\ref{jc2}) gives
\begin{eqnarray}
\label{DS2}
D_{a}(\phi K^{a}_{~b})&=& K^{a}_{~b} D_{a}\phi  - \phi{}^{(5)}R_{cd}h^{d}_{~b}n^{c}.
\end{eqnarray}
Contracting (\ref{pierbulk}) with the induced metric $h^{a}_{~b}$ and the normal vector $n^{a}$ one obtains
\begin{eqnarray}
\label{DS3}
{}^{(5)}R_{cd}h^{d}_{~a}n^{c}=- {1\over \phi} h^{d}_{~a}n^{c} \phi_{;cd}~~.
\end{eqnarray}
After substitution of (\ref{DS3}) to (\ref{DS2}) one gets
 \begin{eqnarray}
\label{DS4}
D_{a}(\phi K^{a}_{~b})&=& K^{a}_{~b} D_{a}\phi + h^{d}_{~b}n^{c} \phi_{;cd}~~.
\end{eqnarray}
Applying (\ref{deko}) to (\ref{DS4}) we have
\begin{eqnarray}
\label{DS5}
D_{a}(\phi K^{a}_{~b})&=& K^{a}_{~b} D_{a}\phi + D_{b}(n^c \nabla_c \phi)-K^{a}_{~b}D_{a}\phi.
\end{eqnarray}
Substituting (\ref{jc3b}) into (\ref{DS5}), we obtain the relation
\begin{eqnarray}
\label{DS6}
D_{a}(\phi K^{a}_{~b})&=&- {\chi \over 16} D_{b}S
\end{eqnarray}
Now taking the covariant derivative of the right hand side of the equation (\ref{jc3c}) multiplied by $\phi$, we have
\begin{eqnarray}
\label{DS7}
D_{a}\left\{ \frac{\chi}{4} \left(S^{a}_{~b}-\frac{1}{4} h^{a}_{~b} S \right) \right\}=
\frac{\chi}{4}  D_{a} S^{a}_{~b} - \frac{\chi}{16} D_{a} S
\end{eqnarray}
Comparison of (\ref{DS6}) with (\ref{DS7}) gives the conservation law of the brane energy-momentum tensor
\begin{eqnarray}
\label{DS8}
D_{a} S^{a}_{~b}=0~~,
\end{eqnarray}
as required.


\begin{thebibliography}{99}

\bibitem{RS} L. Randall and R. Sundrum, Phys. Rev. Lett., {\textbf 83}, 3370 (1999); L. Randall and R. Sundrum, ibidem, {\textbf 83}, 4690 (1999).

\bibitem{brane1} M. Visser, Phys. Lett. B \textbf{159}, 22 (1985);
N. Arkani-Hamed, S. Dimopoulos, and G. Dvali,
Phys. Lett. B \textbf{516}, 70 (1998); I. Antoniadis, N. Arkani-Hamed, S. Dimopoulos, G. Dvali, Phys. Lett. B \textbf{436}, 257 (1998); N. Arkani-Hamed, S. Dimopoulos, and G. Dvali,
Phys. Rev. D \textbf{59}, 086004 (1999).

\bibitem{brane2} P. Bin\'etruy, C. Deffayet and D. Langlois, Nucl.
Phys. B{\bf 565}, 269 (2000); P. Bin\'etruy, C. Deffayet and D. Langlois, Phys.
Lett. B{\bf 477}, 285 (2000); M. Sasaki, T. Shiromizu and K.I. Maeda, Phys.
Rev. D{\bf 62}, 024008 (2000); T. Shiromizu, K.I. Maeda, and M. Sasaki, Phys. Rev. D{\bf 62}, 024012
(2000); S. Mukohyama, T. Shiromizu, and K.I. Maeda,
Phys. Rev. D{\bf 62}, 024028 (2000); A.N. Aliev and A.E. G\"umr\"uk\c{c}\"u\~glu,
Class. Quantum Grav. {\bf 21}, 5081 (2004)

\bibitem{f(R)} A.A. Starobinsky, Phys. Lett. B \textbf{91}, 99 (1980);
G. Magnano and L.M. Soko{\l }owski, Phys. Rev. D \textbf{50}, 5039 (1994);
S. NOjiri and S.D. Odintsov, Phys. Rev. D {\bf 68}, 123512 (2003);
G.J. Olmo, Phys. Rev. Lett. \textbf{98}, 061101 (2007);
S. Capozziello, V.F. Cardone, and A. Troisi, Phys. Rev. D \textbf{71}, 043503 (2005);
S. Capozziello, S. Nojiri, S.D. Odintsov, and A. Troisi, Phys. Lett. B \textbf{639}, 135 (2006);
L. Amendola, D. Polarski, and S. Tsujikawa, Phys. Rev. Lett. \textbf{98}, 131302 (2007);
S. Nojiri and S.D. Odintsov, Phys. Rev. D {\bf 74}, 086005 (2006); S. Nojiri and S.D. Odintsov,
arXiv: 0807.0685.

\bibitem{faraoni10} T. Sotiriou and V. Faraoni, Rev. Mod. Phys. {\bf 82}, 451 (2010), arXiv:0805.1726.

\bibitem{NOreview} S. Nojiri and S.D. Odintsov, Int. J. Geom. Meth. Mod. Phys. {\bf 4}, 115 (2007).

\bibitem{GBbrane} N. Deruelle and T. Dole\v{z}el, Phys. Rev.
D{\bf 62}, 103502 (2000); C. Charmousis, J.F. Dufaux, Class. Quantum Grav. {\bf 19}, 4671 (2002);
S.C. Davis, Phys. Rev. D {\bf 67}, 024030 (2003); 
J.F. Dufaux, J.E. Lidsey, R. Maartens, and M. Sami, Phys. Rev. D {\bf 70}, 083525 (2004);
K.I. Maeda and T. Torii, Phys. Rev. D {\bf 69}, 024002 (2004);
A.N. Aliev, H. Cebeci and T. Dereli, Class. Quantum Grav. {\bf 23}, 591 (2006); 
H. Maeda, V. Sahni, Yu. Shtanov, Phys. Rev. D {\bf 76}, 104028 (2007);
P.S. Apostopoulos {\it et al.}, Phys. Rev. D {\bf 76}, 084029 (2007). 

\bibitem{meissner01} K.A. Meissner and M. Olechowski, Phys. Rev.
Lett. {\bf 86}, 3708 (2001).

\bibitem{paper1} A. Balcerzak and M.P. D\c{a}browski, Phys. Rev. D \textbf{77}, 023524 (2008).

\bibitem{paper2}  A. Balcerzak and M.P. D\c{a}browski, J. Cosmol. Astropart. Phys. \textbf{01}, 018 (2009).

\bibitem{borzeszkowski} H.H.V. Borzeszkowski and V.P. Frolov, Ann. Phys. (Leipzig) {\bf 37}, 285 (1980).

\bibitem{hinterb} E. Dyer and K. Hinterbichler, Phys. Rev. D {\bf 79}, 024028 (2009); arXiv: 0809.4033.

\bibitem{NOlinear} S. Nojiri and S.D. Odintsov, JHEP {\bf 0007},
049 (2000); S. Nojiri, S.D. Odintsov and S. Ogushi, Phys. Rev. D
{\bf 65}, 023521 (2001); S. Nojiri and S.D. Odintsov, Gen. Rel. Grav. {\bf 37}, 1419 (2005).

\bibitem{bd} C. Brans R.H. Dicke, Phys. Rev. {\bf 124}, 925 (1961)~.

\bibitem{israel66} W. Israel, Nuovo Cimento B {\bf 44}, 1 (1966).

\bibitem{BDbrane} M.C.B. Abdalla, M.E.X. Guimar$\tilde{a}$es, and J.M. Hoff da Silva, Eur. Phys. J. C{\bf 55}, 337 (2008).

\bibitem{BDbranenoZ2} M.C.B. Abdalla, M.E.X. Guimar$\tilde{a}$es, and J.M. Hoff da Silva, 0811.4609.

\bibitem{deeg} M. Parry, S. Pichler, and D. Deeg, J. Cosmol. Astropart. Phys. 0504, 014 (2005).

\bibitem{deruelle07} N. Deruelle, M. Sasaki, and Y. Sendouda,
Prog. Theor. Phys. \textbf{119}, 237 (2008).

\bibitem{afonso} V.I. Afonso, D. Bazeia, R. Menezes, and A.Yu. Petrov, Phys. Lett. B{\bf 658}, 71 (2007).

\bibitem{lutrell} S.W. Hawking and J.C. Lutrell, Nucl. Phys. B \textbf{247}, 250 (1984).

\bibitem{wald} R. Wald, General Relativity (University of Chicago Press, Chicago, 1984).

\bibitem{gronhervik} ${\O}.$ Gr${\o}$n, S. Hervik, Einstein's General Theory of Relativity with Modern Application
in Cosmology (Springer Verlag, 2007).

\bibitem{mannheim} P.D. Mannheim, {\it Brane Localized Gravity}
(World Scientific, Singapore, 2005).

\end{thebibliography}
\end{document}